\documentclass[
	reprint,
	groupedaddress,
	amsmath,amssymb,
	prb,
	floatfix,
	superscriptaddress,
	twocolumn,
	aps,
]{revtex4-2}

\usepackage{amsmath}
\usepackage{bm}
\usepackage{amsfonts}
\usepackage{amssymb}
\usepackage{amsthm}
\usepackage[caption=false]{subfig}
\usepackage{xcolor}
\usepackage{graphicx}
\usepackage{hyperref}
\usepackage{cleveref}
\usepackage[T1]{fontenc}
\usepackage{libertine}
\usepackage{lineno}
\usepackage{multirow}

\begin{document}

\title{Thermodynamic and electronic properties of rutile Sn$_{1-x}$Ge$_x$O$_2$ alloys from first principles}
\author{Yann L. M{\"u}ller}
\thanks{These authors contributed equally to this work.}
\affiliation{Laboratory of materials design and simulation (MADES), Institute of Materials, \'{E}cole Polytechnique F\'{e}d\'{e}rale de Lausanne}
\author{Alp Umut Kurbay}
\thanks{These authors contributed equally to this work.}
\affiliation{Department of Materials Science and Engineering, University of Michigan, Ann Arbor, Michigan 48109, USA}
\author{Xiao Zhang}
\affiliation{Department of Materials Science and Engineering, University of Michigan, Ann Arbor, Michigan 48109, USA}
\author{Emmanouil Kioupakis}
\email{kioup@umich.edu}
\affiliation{Department of Materials Science and Engineering, University of Michigan, Ann Arbor, Michigan 48109, USA}
\author{Anirudh Raju Natarajan}
\email{anirudh.natarajan@epfl.ch}
\affiliation{Laboratory of materials design and simulation (MADES), Institute of Materials, \'{E}cole Polytechnique F\'{e}d\'{e}rale de Lausanne}
\affiliation{National Centre for Computational Design and Discovery of Novel Materials (MARVEL), \'{E}cole Polytechnique F\'{e}d\'{e}rale de Lausanne}

\begin{abstract}
    Rutile Sn$_{1-x}$Ge$_x$O$_{2}$ alloys are promising materials for high-power electronic applications due to their dopability and tunable ultra-wide band gaps. We use first-principles density functional theory and statistical mechanics to investigate the crystallographic, electronic, and thermodynamic properties of rutile $\text{Sn}_{1-x}\text{Ge}_x\text{O}_2$ alloys. We predict that the lattice parameters follow Vegard's law, while band gaps calculated with the hybrid HSE06 functional exhibit strong bowing, consistent with experiment. We also predict that the disordered phase has a large positive mixing enthalpy and a slight tendency for Ge–Sn clustering, indicated by weakly negative short-range order parameters. This large positive mixing enthalpy produces a miscibility gap with a critical temperature above 2300 K, implying that the high Ge and Sn solubilities observed in thin-film synthesis cannot be explained by the incoherent phase diagram alone. We demonstrate that coherency strain during epitaxial growth substantially alters phase stability. Calculations of the coherent spinodal show significant suppression of the miscibility gap, reducing the critical temperature to $\approx 900$ K. These coherent phase boundaries account for the experimentally observed high solubilities at typical growth temperatures. Our results indicate that coherency strain stabilizes these metastable alloys and enables bandgap engineering in this ultrawide-bandgap material system.
\end{abstract}

\maketitle

Semiconductors with band gaps wider than that of GaN (3.4 eV), known as ultra-wide-band-gap (UWBG) semiconductors, are promising for high-power electronic devices and UV-visible optoelectronics\cite{wong_ultrawide-bandgap_2021}. UWBG materials can tolerate higher electric fields than conventional semiconductors due to the superlinear increase of the critical dielectric breakdown field with increasing band gap. Several UWBG materials are under investigation \cite{wong_ultrawide-bandgap_2021}, including diamond, cubic boron nitride, aluminum-gallium-nitride alloys, and semiconducting oxides such as $\beta\text{--Ga}_2\text{O}_3$, SnO$_2$, rutile GeO$_{2}$, and rutile SiO$_2$\cite{Lyons_2024,chae2025}.

Rutile GeO$_{2}$ (r-GeO$_2$) and its alloys with SnO$_2$, in particular, have recently emerged as promising materials for UWBG applications due to their attractive electronic properties. r-GeO$_2$ has been predicted to exhibit
an ultra-wide band gap of 4.68 eV \cite{10.1063/1.5111318},
efficient n-type (and potential ambipolar) doping \cite{10.1063/1.5088370}, high electron and hole mobilities\cite{10.1063/5.0033284} and high thermal conductivity \cite{10.1063/5.0011358}.
The large band gap and high carrier mobilities of r-GeO$_2$ yield a high Baliga's figure of merit\cite{chae2022} surpassing other power electronic candidates such as Ga$_2$O$_3$.
The n-type dopability has been experimentally demonstrated for bulk \cite{2025_Galazka_Rutile_GeO2} and thin-film \cite{2025_Shimazoe} samples, while MOSFET devices \cite{11271013} and vertical Schottky barrier diodes \cite{2025_Kanegae} based on r-GeO$_2$ have recently been demonstrated.

Alloying r-GeO$_2$ with r-SnO$_2$ yields a tunable band gap ranging from 4.68 eV (r-GeO$_{2}$) to 3.6 eV (r-SnO$_{2}$). Pulsed laser deposition\cite{nagashima_deep_2022,Kluth2024_GeSnO2_Bowing}, chemical vapor deposition\cite{takane_band-gap_2022}, sputtering\cite{10.1116/6.0003960,Abed2025}, and molecular beam epitaxy\cite{10.1063/5.0018031,liu_unraveling_2025} have been utilized to synthesize rutile Sn$_{1-x}$Ge$_x$O$_{2}$ (r-Sn$_{1-x}$Ge$_x$O$_{2}$). However, film quality is sensitive to the composition, often suffering from amorphization and challenges in obtaining single-phase  as the Ge content increases.

In this study, we systematically investigate the thermodynamic and electronic properties of r-$\text{Sn}_{1-x}\text{Ge}_x\text{O}_2$ alloys using density functional theory (DFT) and statistical mechanics-based calculations. Our results reveal that the lattice parameters of disordered solid solutions follow Vegard’s law, and the band gaps exhibit a bowing effect. Finite-temperature thermodynamic calculations, based on both an approximate free-energy model and more accurate Monte-Carlo simulations informed by on-lattice cluster expansions, predict a high-temperature miscibility gap consistent with early experiments. Finally, we show that the variation in solid solubilities observed across synthesis techniques likely stems from the suppression of the miscibility gap due to coherency strain. Our results highlight the importance of accounting for coherency strain effects on phase stability in alloys of r-GeO$_{2}$.

Density functional theory (DFT) calculations with the Perdew-Burke-Ernzerhof (PBE)\cite{perdew_generalized_1996}, r${^2}\mathrm{SCAN}$\cite{furness_accurate_2020}, and HSE06\cite{heyd_hybrid_2003} functionals were used to study the crystallographic, thermodynamic and electronic properties of $\text{r-Sn}_{1-x}\text{Ge}_x\text{O}_2$. We employed HSE06 hybrid-functional calculations, in particular, to obtain accurate band gaps, as the other two functionals systematically underestimate them. All DFT calculations used projector augmented-wave (PAW) potentials as implemented in the Vienna \emph{Ab Initio} Simulation Package (VASP)\cite{VASP, PAW}. We relaxed structures with the PBE\cite{perdew_generalized_1996} and r${^2}\mathrm{SCAN}$ functionals until forces on atoms converged to less than 0.01 eV/\AA. These calculations treated Ge $4s^2 3d^{10} 4p^2$ and Sn $5s^2 4d^{10} 5p^2$ as valence electrons, with a plane-wave energy cutoff of 600 eV and a \(\Gamma\)-centered k-point grid density of 35$\text{\AA}$. For HSE06 calculations, we used GW-compatible PBE pseudopotentials with a mixing parameter of 0.315 for all compositions. The HSE06 calculations included the Ge \(3s^2\,3p^6\) and Sn \(4s^2\,4p^6\) electrons in addition to those above, with a plane-wave energy cutoff of 950 eV to ensure total energy convergence within 0.1 meV/atom. We sampled the Brillouin zone with a $\Gamma$-centered \(4\times 4\times 6\) grid for binary unit cells and a \(\Gamma\)-centered \(2\times 2\times 2\) grid for alloy supercells.

\begin{figure*}[htbp]
    \centering
    \includegraphics[width=0.82\linewidth]{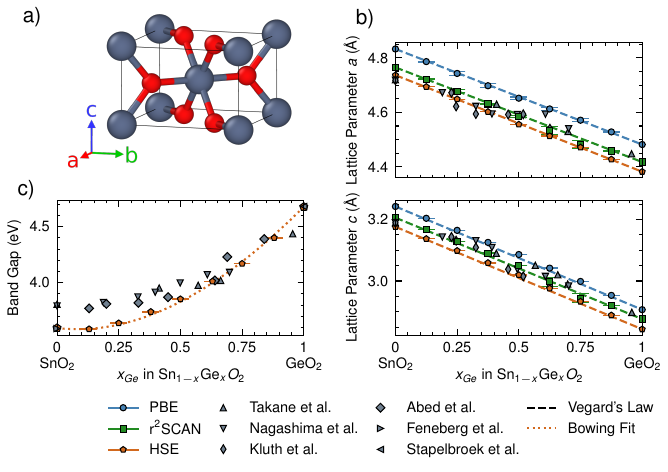}
    \caption{a) Unit cell of the rutile structure. The blue atoms correspond to cation sites occupied by Ge or Sn atoms, while red sites correspond to oxygen atoms. b) Calculated PBE (blue circles), r$^2$SCAN (green squares), HSE (orange pentagons) lattice constants $a$ (top) and $c$ (bottom), for $\text{Sn}_{1-x}\text{Ge}_x\text{O}_2$ alloys at x = 0, 0.125, 0.25, 0.375, 0.5, 0.625, 0.75, 0.875, 1. The values for the alloys are calculated by averaging three SQS supercells, each containing 72 atoms (a $2\times2\times3$ supercell of the primitive unit cell). The error bars represent the standard deviation. c) Calculated HSE bandgaps (orange pentagons) for $\text{Sn}_{1-x}\text{Ge}_x\text{O}_2$ alloys using 72 atom SQSs. The bowing fit is determined using a second-order composition-dependent model. The experimental data for alloy thin films are given by \cite{takane_band-gap_2022} (upward triangles), \cite{nagashima_deep_2022} (downward triangles)\label{fig:structure}, \cite{Abed2025} (diamonds), and \cite{Kluth2024_GeSnO2_Bowing} (rhombuses), while the data for the bulk crystals is given by \cite{Feneberg2014_SnO2_Dielectric} (rightward triangles) and \cite{StapelbroekEvans1978_GeO2_UVabsorption} (leftward triangles).
    }
\end{figure*}

The rutile structure, shown in \cref{fig:structure}a), accommodates disorder on the cation sublattice, where each site can be occupied by either Ge or Sn atoms. We do not consider disorder on the anion (oxygen) sublattice. We approximate properties of the disordered phase using special quasi-random structures (SQS)\cite{PhysRevLett.65.353}. SQS-based chemical orderings were generated in a simulation cell containing 72 atoms (24 formula units) for seven equally spaced compositions ranging from pure SnO$_{2}$ ($x=0$) to GeO$_{2}$ ($x=1$), ensuring that pair cluster functions were disordered up to the sixth nearest neighbor shell. We relaxed structures with respect to lattice parameters and atomic positions using the PBE, r$^{2}$SCAN, and HSE06 functionals to obtain lattice parameters and total energies. The band gaps of the SQS structures were calculated with the HSE06 functional.

Understanding the finite-temperature synthesizability of disordered $\text{r-Sn}_{1-x}\text{Ge}_x\text{O}_2$ alloys requires quantifying the thermodynamics of the binary alloy. Finite-temperature free energies can be estimated through several methods. The simplest approximation uses the energies of SQS structures to construct a free energy model. The formation energy ($\Delta e_{f}$) for a specific arrangement of Ge and Sn on the rutile lattice is given by
\begin{widetext}
    \begin{equation}
        \label{eq:formation_energy}
        \Delta e_{f}(\text{Sn}_{2(1-x)}\text{Ge}_{2x}\text{O}_4) = \frac{2}{M}\left(E(\text{Sn}_{(M-N)}\text{Ge}_{N}\text{O}_{2M}) - N E(\text{Ge} \text{O}_{2}) - (M-N) E(\text{Sn} \text{O}_{2})\right)
    \end{equation}
\end{widetext}
where $E(\text{Sn}_{(M-N)}\text{Ge}_{N}\text{O}_{2M})$ is the total energy of a supercell containing $M$ total cation sites. $N$ of these sites are occupied by Ge (giving a composition $x=N/M$). $E(\text{Ge} \text{O}_{2})$ and $E(\text{Sn} \text{O}_{2})$ are the total energies per formula unit of bulk rutile GeO$_{2}$ and SnO$_{2}$, respectively. \Cref{eq:formation_energy} is the formation energy per rutile primitive cell containing two formula units of r-Sn$_{1-x}$Ge$_x$O$_{2}$. The formation energies of the SQS structures are used to fit the mixing enthalpy ($\Delta h_{dis}$) of the disordered solid solution with a subregular solution model
\begin{equation}
    \Delta h_{dis}(\text{Sn}_{2(1-x)}\text{Ge}_{2x}\text{O}_4) = \alpha x(1-x) + \beta x(1-x)(1-2x)
    \label{eq:H}
\end{equation}
where $\alpha$ and $\beta$ are parameters fitted to the DFT-computed SQS formation energies.
The ideal mixing entropy ($\Delta s_{mix}$) per primitive cell is estimated as
\begin{equation}
    \Delta s_{dis}(\text{Sn}_{2(1-x)}\text{Ge}_{2x}\text{O}_4) = -2 k_B \big(x \ln(x) + (1-x)\ln(1-x) \big)
    \label{eq:S}
\end{equation}
where the factor of 2 arises from the two cation sites within the rutile primitive unit cell. The free energy of the disordered phase is computed as $\Delta g_{\textrm{disordered}}(\text{Sn}_{2(1-x)}\text{Ge}_{2x}\text{O}_4, T) = \Delta h_{dis} - T \Delta s_{dis}$.

The approximate free energy model constructed with the formation energies of the SQS structures and ideal solution mixing entropy does not account for the formation of any ordered phases or the formation of short-range order within the disordered solid solution. A more accurate estimate of finite-temperature thermodynamics in systems with order-disorder transitions can be obtained from statistical mechanics simulations. These simulations are informed by surrogate models, such as on-lattice cluster expansions (CE), which are parameterized using DFT calculations for a small set of symmetrically distinct arrangements on a parent crystal structure. Cluster expansion Hamiltonians\cite{sanchez_generalized_1984, fontaine_cluster_1994} are effective tools for assessing the thermodynamic properties of crystalline materials\cite{lee2026,muller2024,muller2025,natarajan2017a,natarajan2018a,paetsch2026, angqvist_optimization_2016, angqvist_understanding_2017, brorsson_firstprinciples_2021, lee_thermodynamics_2024, wang_kinetic_2023}. Within the cluster expansion formalism, any arrangement of Sn and Ge on the rutile lattice is represented by an occupation vector $\vec{\sigma} = [\sigma_1 \; \sigma_2 \; \ldots \sigma_N]^T$. The $i$th entry $\sigma_i$ takes the value $-1$ if the $i$th lattice site is occupied by Sn and $1$ if occupied by Ge. The formation energy $\Delta e_{f}(\vec{\sigma})$ of an arbitrary arrangement $\vec{\sigma}$ in this alloy is given by:
\begin{equation}
    \Delta e_{f}(\vec{\sigma}) = \frac{1}{M}\left(V_0 + \sum_i V_i \sigma_i + \sum_{i,j} V_{ij} \sigma_i \sigma_j + \ldots\right)
\end{equation}
where $V_0$, $V_i$, and $V_{ij}$ are fitting coefficients known as effective cluster interactions (ECI). The CE is then used within Monte-Carlo simulations and free-energy integration methods to compute the finite-temperature free energy of all relevant phases in the alloy. Common tangent constructions are then employed to compute phase boundaries.

\Cref{fig:structure} shows the calculated lattice constants and band gap of r-Sn$_{1-x}$Ge$_x$O$_2$ alloys as a function of alloy composition $x$. \Cref{fig:structure}b shows the calculated PBE, r$^2$SCAN, and HSE lattice constants averaged over three SQS supercells containing 72 atoms. The HSE and r$^2$SCAN results are in excellent agreement with experimental studies, whereas PBE overestimates the lattice constants due to the well-known underbinding tendency of the functional. The calculated lattice constants clearly demonstrate the validity of Vegard’s law in all three cases.

The bandgaps for the SQS structures computed with the HSE06 hybrid functional are shown in \cref{fig:structure}b. Bandgaps are averaged over three different SQSs at each composition. We find a direct bandgap at the $\Gamma$ point across all compositions, which is fit to a second-order composition-dependent bowing model:
\begin{equation}
    E_g(x) = xE_{g,\mathrm{GeO}_2}
    + (1-x)E_{g,\mathrm{SnO}_2}
    - x(1-x)\big[b\,x + c(1-x)\big]
\end{equation}

The second-order bowing fit is employed here to more effectively capture variations of the bowing parameter with respect to composition, as observed in other UWBG semiconductor alloys such as Al$_{1-x}$In$_{x}$N \cite{Schulz2013_AlInN_Bowing}.
This fit yields bowing parameters $b = 1.115 \pm 0.01$ eV  and $c = 1.336 \pm 0.005$ eV, indicating stronger bowing at lower Ge content.
Both of these values are in agreement with experimental results reported by Takane\textit{~et~al.}, who found a bowing parameter of 1.2~eV using a first-order fit \cite{takane_band-gap_2022}.
It should be noted that while our calculated bandgap of 3.6~eV for SnO$_2$ agrees with experimental studies of the bulk material, this value is roughly 200~meV lower than that reported in experimental optical absorption measurements for thin-film samples \cite{takane_band-gap_2022, nagashima_deep_2022}.
This discrepancy may arise from the dipole-allowed or -forbidden nature of the transitions depending on symmetry breaking due to alloy disorder, or from strain effects in the thin-film samples used experimentally.

\begin{figure*}[htbp]
    \centering
    \includegraphics[width=1\linewidth]{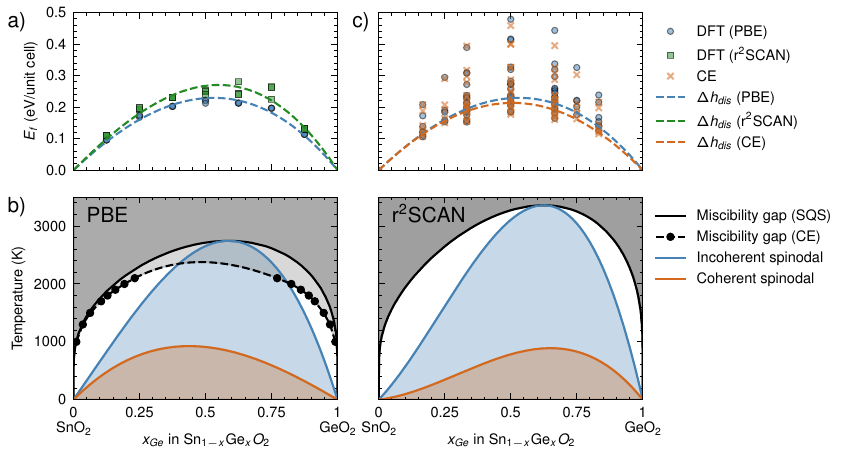}
    \caption{a) DFT formation energies of the SQS structures evaluated with PBE (blue circles) and r$^{2}$SCAN (green squares) functionals. The dashed lines represent the enthalpy of the disordered phase estimated with the subregular solution model.
        b) Phase diagrams of the r-$\text{Sn}_{1-x}\text{Ge}_x\text{O}_2$ computed using PBE functional (left) and r²SCAN functional (right). The black solid line indicates the miscibility gap, the blue line the incoherent spinodal, and the orange line the coherent spinodal along the $[001]$ direction. The miscibility gap obtained through free energy integration using the CE is indicated by black dots in the PBE phase diagram.
        c) Convex hull computed with 80 symmetrically distinct orderings on $\text{r-Sn}_{1-x}\text{Ge}_x\text{O}_2$ predicted by DFT (blue circles) and CE (orange crosses), using the PBE functional. The blue dashed lines are the disordered enthalpy estimated with the subregular solution model fit to the DFT formation energies of the SQS structures. The orange dashed line is the enthalpy of a perfectly disordered solid solution estimated with the CE model.}
    \label{fig:phase_diagram}
\end{figure*}

\Cref{fig:phase_diagram} compares the 0 K and finite-temperature phase stability of r-Sn$_{1-x}$Ge$_{x}$O$_{2}$ alloys as computed with different exchange-correlation functionals. The formation energies of the 21 SQS orderings, shown in \cref{fig:phase_diagram}a, are qualitatively similar when computed with PBE and r$^{2}$SCAN. However, the r$^{2}$SCAN values are up to $17\%$ higher than the PBE values, a difference that is largest at high germanium compositions. Both DFT functionals predict positive mixing enthalpies. This result indicates low-temperature phase separation and suggests that mixing between the terminal oxides occurs only at higher temperatures. We used the SQS formation energies of \cref{fig:phase_diagram}a to fit the subregular solution model for the disordered phase enthalpy (\cref{eq:H}). The resulting fitting parameters are listed in \cref{tab:params}. The large positive values of $\alpha$ in \cref{tab:params} indicate significant immiscibility in this alloy.

\Cref{fig:phase_diagram}b shows the finite-temperature phase diagrams computed with the PBE and r$^{2}$SCAN functionals using the SQS-based free energy model. Both phase diagrams predict a wide miscibility gap, indicating phase separation. For an equiatomic solid solution, the critical temperature is predicted to be $\sim$2750 K with PBE and $\sim$3350 K with r$^{2}$SCAN. The higher critical temperature from r$^{2}$SCAN is a direct result of the larger SQS formation energies predicted by that functional. These results are in qualitative agreement with the experimental phase diagram reported by Watanabe \emph{et al.}\cite{watanabe1983solid}. Their work suggests $\approx$4\% solubility of GeO$_{2}$ in SnO$_{2}$ at 1250 \textdegree C, with very little solubility of SnO$_{2}$ in GeO$_{2}$. Our calculations show a similar trend, predicting low solubility of GeO$_{2}$ in SnO$_{2}$ and minimal dissolution at the GeO$_{2}$-rich end of the phase diagram.

\begin{table}
    \centering
    \begin{tabular}{c|c c}
                                       & $\alpha$ (eV/unit cell) & $\beta$ (eV/unit cell) \\
        \hline
        $\Delta h_{dis}$ (PBE)         & 0.915                   & -0.117                 \\
        $\Delta h_{dis}$ (r$^{2}$SCAN) & 1.071                   & -0.219                 \\
    \end{tabular}
    \caption{Fitting parameters for the subregular solution model of the disordered phase enthalpy, based on \cref{eq:H}.}
    \label{tab:params}
\end{table}

The SQS-based approximate free energy model provides finite-temperature predictions consistent with experiment, but quantitative differences in phase stability can arise from short or long-range ordering effects. A rigorous evaluation of finite-temperature phase stability was performed using the CE method. We computed the formation energies of all 80 symmetrically distinct chemical decorations of Ge and Sn on the rutile cation sublattice within supercells containing up to 6 cation sites using the PBE functional. The computed formation energies are shown in \cref{fig:phase_diagram}c. No ordered ground states are predicted to be stable at 0 K. The positive formation energies suggest immiscibility at low temperatures, in agreement with the formation energies of the SQS orderings in \cref{fig:phase_diagram}a. To parameterize the CE model based on these 80 formation energies, descriptors of chemical ordering were computed with the \emph{Clusters Approach to Statistical Mechanics (CASM)} software package\cite{puchala_casm_2023, van_der_ven_first-principles_2018, puchala_thermodynamics_2013} for clusters containing up to 4 sites located within a maximum distance of 9 \AA. A 5-fold cross-validated Lasso model\cite{tibshirani_regression_1996}, as implemented in the \texttt{sklearn} package\cite{pedregosa_scikit-learn_nodate}, was utilized for the linear regression. The resulting CE model achieved a root mean squared error (RMSE) of 9 meV/unit cell (1.5 meV/atom). A comparison between the DFT and CE formation energies is shown in \cref{fig:phase_diagram}c.

\Cref{fig:phase_diagram}b compares the finite-temperature phase diagram for r-Sn$_{1-x}$Ge$_x$O$_{2}$ derived from rigorous CE-based statistical mechanics with the prediction from the SQS-based approximate free energy model. The CE-based phase boundaries were determined using semi-grand canonical Monte Carlo simulations, followed by free energy integration and a common tangent construction. The resulting miscibility gap is shown by the black dots in \cref{fig:phase_diagram}b (left). The critical temperature predicted by the CE model is approximately $400$K lower than that predicted by the PBE-based subregular model.

Small discrepancies between the phase boundaries predicted by the CE and SQS-based models arise from differences in the mixing enthalpy and short-range order (SRO) effects. \Cref{fig:phase_diagram}c compares the CE-predicted enthalpy of the disordered solid solution to the SQS-informed subregular solution enthalpy. The CE mixing enthalpy is approximately $7\%$ lower than the SQS-based PBE value. This enthalpy difference alone accounts for a $\approx$100 K reduction in the predicted critical temperature.

\begin{figure}[!h]
    \centering
    \includegraphics[width=1\linewidth]{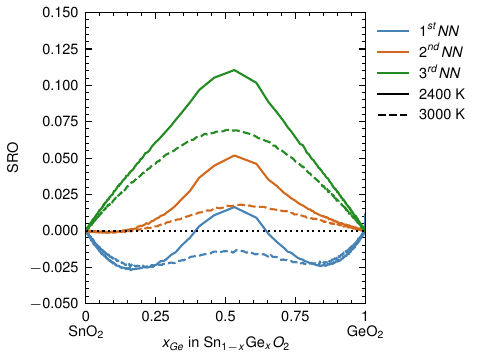}
    \caption{Plot of the Ge-Sn SRO as a function of the composition for the first (blue), second (orange), and third (green) nearest-neighbor shells, computed at $2400$ K (solid line) and at $3000$ K (dashed line).}
    \label{fig:sro}
\end{figure}

The remaining discrepancy arises from SRO near the phase transition. SRO formation lowers both the enthalpy and entropy of the disordered state compared to a perfectly random solution. When the enthalpy gain offsets the entropy penalty, the free energy of the disordered state decreases relative to the perfectly disordered random solution. \Cref{fig:sro} displays the SRO of Ge-Sn pairs in the first, second, and third nearest-neighbor shells obtained from Monte Carlo simulations at 2400 K and 3000 K. A negative SRO parameter indicates more Ge-Sn pairs relative to a random alloy, while a positive value signifies fewer pairs. \Cref{fig:sro} reveals a slight tendency for the clustering of Ge-Sn pairs in the first-nearest-neighbor shell (vertical edges in \cref{fig:structure}a). An elevated number of Ge-Ge and Sn-Sn pairs are predicted to occur in the second and third nearest-neighbor shells. The degree of SRO is small, with absolute values remaining below 0.15. \Cref{fig:sro} shows that SRO becomes more positive as the temperature decreases. This is consistent with a phase-separating behavior at low temperatures. The emergence of SRO near the phase transition likely accounts for the lower critical temperature predicted by the CE method relative to the approximate free energy model. The similarities in PBE and r$^{2}$SCAN formation energies suggest that a similar discrepancy would emerge if the CE method were applied to the r$^{2}$SCAN data. Nevertheless, the overall analysis presented in \cref{fig:phase_diagram} indicates a critical temperature for the disordered solid solution of $\approx$2350 - 2750 K (PBE) or $\approx$3350 K (r$^2$SCAN).

Our finite-temperature phase diagrams agree well with early experiments for bulk materials\cite{watanabe1983solid}. However, they appear to contradict recent reports of wide-ranging miscibility in thin films\cite{liu_unraveling_2025,takane_band-gap_2022} and a recent computational prediction\cite{liu_unraveling_2025}. On the computational side, the critical temperature predicted by our thermodynamic model is approximately 3 times higher than the calculated value of $\approx 986$ K reported by Liu \textit{et al.}\cite{liu_unraveling_2025}. This discrepancy exists despite the use of identical DFT functionals in both studies. The difference likely stems from the mixing enthalpy reported by Liu \textit{et al.}\cite{liu_unraveling_2025}, which is three times lower than our calculated values (\cref{tab:params}). The significant difference in free energies leads to qualitatively different behavior. Liu \textit{et al.} attribute their ability to grow films with 34\% Ge in SnO$_{2}$ at 600 \textdegree C to the metastable solubility limit predicted by their calculated spinodal. In contrast, the spinodal predicted by our models (shown by the blue curves in \cref{fig:phase_diagram}) indicates a metastable solubility limit of only $\approx$10\% at 600 \textdegree C. The observed higher solubility of Ge in SnO$_{2}$ is unlikely to be explained solely by the incoherent spinodal decomposition and miscibility gap shown in \cref{fig:phase_diagram}.

The phase diagrams in \cref{fig:phase_diagram} represent incoherent equilibria, showing phase boundaries and solubility limits for phases that form without microstructural constraints. In epitaxially grown thin films, incoherent equilibria are often not achieved. Instead, constraints imposed by the growth conditions force coexisting phases to share the same lattice parameters, leading to coherent phase coexistence. Coherent phase diagrams are known to differ quantitatively from their incoherent counterparts. This coherency can significantly alter solubility limits and phase stability in materials classes ranging from metallic alloys\cite{NATARAJAN2016367} to thermoelectric\cite{doak_coherent_2012} and battery materials\cite{Malik_2013}.

Precisely estimating coherent phase equilibrium is challenging as phase compositions are sensitive to microstructure, local stresses, and composition-dependent material properties. Qualitative estimates, however, can be obtained using approximations introduced by Cahn\cite{cahn_spinodal_1961} and implemented by Doak and Wolverton\cite{doak_coherent_2012}. Within this framework, the enthalpy of the disordered phase under the constraint of coherent equilibrium is:
\begin{equation}
    \label{eq:coherent_mixing_enthalpy}
    \Delta h^{coh}_{dis}(x) = \Delta h_{dis}(x) - \Delta e_{cs}(x)
\end{equation}
where $\Delta h_{dis}$ is the incoherent enthalpy of the disordered state (\cref{eq:H}) and $\Delta e_{cs}$ is the coherency strain energy due to phase coexistence at the same lattice parameters. As described by Doak and Wolverton\cite{doak_coherent_2012}, an approximate coherency strain energy can be computed along different coexistence planes. This calculation involves fixing the in-plane lattice parameters of the terminal oxides and minimizing the energy with respect to the out-of-plane lattice parameter. The total coherency strain energy for coexistence is the composition-weighted sum of the strain energies of the terminal oxides. The coherency strain energy used in \cref{eq:coherent_mixing_enthalpy} is the minimum value of this composition-weighted energy found across all possible coexistence directions.

The spinodal line, which dictates the metastable limit of Ge or Sn solubility, can be computed under coherent coexistence by finding the locus of points where the second derivative of the coherent disordered free energy is zero. This free energy is given by $\Delta g_{dis}^{coh} = \Delta h^{coh}_{dis}(x) - T \Delta s_{dis}(x)$, where the ideal solution entropy is used for $\Delta s_{dis}(x)$. The coherent enthalpy (\cref{eq:coherent_mixing_enthalpy}) was fitted to a subregular solution model, and the spinodal lines for coherent phase equilibria were then computed analytically using the methodology outlined by Doak and Wolverton\cite{doak_coherent_2012}.

\begin{table}[]
    \centering
    \begin{tabular}{c|c c}
                                            & $\alpha$ & $\beta$ \\
        \hline
        $\Delta e_{cs}^{001}$ (PBE)         & 0.601    & -0.144  \\
        $\Delta e_{cs}^{001}$ (r$^{2}$SCAN) & 0.801    & -0.144  \\
        $\Delta e_{cs}^{100}$ (PBE)         & 1.176    & -0.001  \\
        $\Delta e_{cs}^{100}$ (r$^{2}$SCAN) & 1.331    & 0.027   \\
        $\Delta e_{cs}^{101}$ (PBE)         & 1.833    & -0.361  \\
        $\Delta e_{cs}^{101}$ (r$^{2}$SCAN) & 2.108    & -0.391
    \end{tabular}
    \caption{Fitting parameters for the subregular solution model of the coherency strain energy, based on \cref{eq:H}.}
    \label{tab:coherent_params}
\end{table}

The coherent spinodals computed assuming coexistence along the $(001)$ plane are shown in \cref{fig:phase_diagram}b. The corresponding model parameters are listed in \cref{tab:coherent_params}. The resulting coherent spinodal, shown by the orange lines in \cref{fig:phase_diagram}b, is significantly suppressed compared to the incoherent spinodal. The predicted coherent spinodal reaches a maximum temperature of $930$ K (at $\approx 44\%$ Ge) with PBE and $890$ K (at $\approx 65\%$ Ge) with r$^{2}$SCAN. Although the incoherent mixing enthalpies are more positive with r$^{2}$SCAN than with PBE, the coherent spinodal temperatures predicted by both functionals are very similar. This arises due to the stiffer elastic constants predicted by r$^{2}$SCAN.

The metastable solubility limits predicted by the coherent spinodals in \cref{fig:phase_diagram} align well with recent experimental efforts to synthesize r-Sn$_{1-x}$Ge$_x$O$_{2}$. For instance, at 600 \textdegree C, the predicted coherent solubility limit of Ge ($x \approx 0.3$) closely matches the experiments by Liu \emph{et al.}\cite{liu_unraveling_2025}. In the experiments by Takane \emph{et al.}\cite{takane_band-gap_2022}, a higher growth temperature of 725 \textdegree C was used. At this temperature, our calculations predict nearly complete solid solubility across the entire composition space, consistent with their synthesis of alloys across the full composition range.

Our computational predictions for the crystallographic parameters and finite-temperature thermodynamics of r-Sn$_{1-x}$Ge$_x$O$_{2}$ show qualitative agreement with experiments\cite{liu_unraveling_2025,takane_band-gap_2022}. The coherent spinodals predicted in \cref{fig:phase_diagram} assume that entropy arises solely from configurational disorder, that the coherency strain energy can be approximated from the strain energies of the pure phases, and that the lattice is defect-free. More rigorous estimates would require advanced techniques such as machine-learned interatomic potentials or CE models that couple ordering and strain degrees of freedom\cite{behara_PhysRevMaterials.8.033801}. Such atomistic models would enable rigorous treatment of vibrational contributions, coherency strain energies and defects such as dislocations or semi-coherent interfaces, which have been shown to affect coherent phase separation\cite{leonard1998,hu2001,hu2002}. Furthermore, during thin-film growth the substrate dictates the film orientation. The actual coexistence plane and lattice parameters may differ from our estimates, yielding different metastability limits. For instance, our estimates suggest that enforcing coexistence along the $(100)$ or $(101)$ planes could produce complete miscibility even at low temperatures due to large coherency strain energies. If growth conditions force coexistence along these high-strain planes, interfacial defects would likely form to reduce the total free energy. Finally, our model does not account for amorphization, which has been observed experimentally. Predicting amorphization would require estimates of the amorphous phase free energy to compare against that of coherent coexistence during synthesis.

We investigated the crystallographic, electronic, and thermodynamic properties of the rutile $\text{Sn}_{1-x}\text{Ge}_x\text{O}_2$ alloy using density functional theory at varying levels of accuracy. Lattice parameters predicted by the PBE, r$^{2}$SCAN, and HSE06 functionals agree well with experimental values and follow Vegard's law, consistent with previous reports\cite{takane_band-gap_2022, liu_unraveling_2025}. Band gap calculations using the HSE06 functional reveal strong bowing at low Ge content, also consistent with previous observations\cite{takane_band-gap_2022}. Finite-temperature phase stability calculations indicate a slightly negative SRO parameter for Ge–Sn pairs in the first neighbor shell and show that the incoherent miscibility gap cannot explain the high Ge solubilities observed in recent experiments, as it predicts low solubility at typical synthesis temperatures. Our calculations instead reveal that coherency strain suppresses the miscibility gap, enabling large Ge and Sn solubilities in agreement with experiments. These findings provide guidance for engineering UWBG semiconductors with tunable properties and highlight the role of coherency strain in enabling bandgap engineering.

YLM and ARN acknowledge funding from the NCCR MARVEL, a National Centre of Competence in Research supported by the Swiss National Science Foundation (grant 205602), which also provided hospitality and support for EK's sabbatical stay at EPFL, during which part of this work was performed. AUK, XZ, and EK were supported by the National Science Foundation through grant 2328701 which is supported in part by federal agency and industry partners through the Future of Semiconductors (FuSe) program. Computational resources were provided by Anvil at Purdue University through allocation DMR200031 from the ACCESS program, supported by U.S. National Science Foundation grants 2138259, 2138286, 2138307, 2137603, and 2138296, and by Eiger at the Swiss National Supercomputing Centre (project ID mr30) under the NCCR MARVEL allocation.

\end{document}